\documentstyle[twocolumn,floats,ifthen,aps,prl]{revtex}

\begin{document} 
 
\draft 
%\preprint{draft} 

\title{ 
Continuous quantum measurement with particular output: 
pure wavefunction evolution instead of decoherence 
}

\author{Alexander N. Korotkov}
\address{ 
GPEC, Departement de Physique, Facult\'e des Sciences de Luminy, 
Universit\'e de la M\'editerran\'ee, 
13288 Marseille, France \\ 
and Nuclear Physics Institute, Moscow State University,
Moscow 119899, Russia}
 
\date{\today} 
 
\maketitle 
 
\begin{abstract} 
        We consider a continuous measurement of a two-level 
system (double-dot) by weakly coupled detector (tunnel point contact 
nearby). While usual treatment leads to the gradual system 
decoherence  due to the measurement, we show 
that the knowledge of the measurement result can restore the pure
wavefunction at any time (this can be experimentally verified). 
The formalism allows to 
write a simple Langevin equation for the random evolution of the system
density matrix which is reflected and caused by the stochastic 
detector output. Gradual wavefunction ``collapse'' 
and quantum Zeno effect are naturally described by the equation.
\end{abstract} 
 
 
\narrowtext 
 
%\vspace{1ex} 

        The problem of quantum measurements has a long history,
however, it still attracts considerable attention and causes
discussions and even some controversy, mainly about the wavefunction
``collapse'' postulate (see, e.g., Ref.\ \cite{Wheeler}).  
        Among different modern approaches to this problem let us 
mention the idea to replace the collapse postulate by the gradual 
decoherence of the density matrix due to the interaction with the 
detector \cite{Zurek} and 
the approach of a stochastic evolution of the wavefunction
(see, e.g., \cite{Gisin,Carmichael,Plenio}). 
The renewed interest to the measurement problem 
is justified by the development of the experimental technique,
which allows more and more experimental studies of the quantum
measurement in optics and mesoscopic structures.  
The problem has also the 
close connection to the rapidly growing fields of quantum cryptography 
and quantum computing (see, e.g., \cite{q-comp}).

        In the recent experiment \cite{Buks} with  ``which-path''
interferometer the suppression of Aharonov-Bohm interference 
due to the detection of which path an electron chooses, was observed. 
The weakly coupled quantum point contact was used as a detector.
The interference suppression in this experiment can be quantitatively 
described by the decoherence due to the measurement process 
\cite{Gurvitz,Aleiner,Levinson,Stodolsky}.  

We will consider somewhat different setup: 
two quantum dots occupied by one electron and a weakly coupled 
detector (point contact nearby) measuring the position of the electron.
The decoherence of the double-dot density matrix  due to the 
measurement has been analyzed for this setup in Refs.\ 
\cite{Gurvitz,Stodolsky}. 
        In the present letter we answer the following questions: 
how the detector current looks like (as a function of time) 
and what is the proper double-dot density matrix 
for particular detector output. 
We show that the models of point contact considered in Refs.
\cite{Gurvitz,Aleiner,Levinson} describe an ideal detector.  
In this case the density matrix decoherence is just a 
consequence of ignoring the measurement result. 
The observer who follows the detector output knows 
the pure wavefunction at each moment of time. 
Moreover, a ``mixed'' density matrix can be gradually purified 
during the measurement.

        Similar to Ref.\ \cite{Gurvitz} let us describe
the double-dot system and the measuring point contact by the Hamiltonian
       $ {\cal H}={\cal H}_{DD}+{\cal H}_{PC}+{\cal H}_{int}$ ,
where 
${\cal H}_{DD} = (\epsilon /2) (c_1^\dagger c_1-c_2^\dagger c_2)+ 
H (c_1^\dagger c_2+ c_2^\dagger c_1) $
is the Hamiltonian of the double-dot,   
${\cal H}_{PC}=\sum_l E_la_l^\dagger a_l +\sum_r E_r a_r^\dagger a_r +
\sum_{l,r} T (a_r^\dagger a_l+a_l^\dagger a_r) $
describes the tunneling through the point contact ($H$ and $T$ are real), 
and ${\cal H}_{int}= \sum_{l,r} \Delta T \, c_2^\dagger c_2 
(a_r^\dagger a_l + a_l^\dagger a_r)$,
i.e.\ the tunneling matrix element for the point contact is $T$ or 
$T+\Delta T$ depending on which dot is occupied.
So, the average current $I_1=2\pi T^2 \rho_l \rho_r e^2V/\hbar$ flows 
through the detector when the electron is in the first dot
($V$ is voltage), while the current is
$I_2=I_1+\Delta I=2\pi (T+\Delta T)^2\rho_l\rho_r e^2V/\hbar$ 
when the second dot is occupied.

        We make an important assumption of weak coupling between
the double-dot and the detector (the better term would be 
``weakly responding'' detector),
        \begin{equation}
| \Delta I | \ll I_0= (I_1 +I_2)/2,
        \label{weak}\end{equation}
so that many electrons, $N \agt (I_0/\Delta I)^2$, should pass through 
the point contact before
the observer is able to distinguish which dot is occupied.
This assumption allows the classical description 
of the detector, namely the coherence between the 
quantum states with different number of electrons passed through
the detector can be neglected \cite{bigDI}.

        The decoherence rate $\Gamma_d = (\sqrt{I_1/e}-\sqrt{I_2/e})^2/2$
of the double-dot density matrix ${\bf \sigma} (t)$ due to 
the measurement by point
contact has been obtained in Ref.\ \cite{Gurvitz}.
In the weakly-coupled limit (\ref{weak}) it can be replaced
by $\Gamma_d= (\Delta I)^2/8eI_0$ or by expression 
        \begin{equation}
\Gamma_d = (\Delta I)^2/4S_I,
        \label{gam_d-m}\end{equation} 
where $S_I=2eI_0$ is the usual Schottky formula for the shot noise
spectral density $S_I$. 
Equation (\ref{gam_d-m}) has been also obtained in Refs.\
\cite{Aleiner,Levinson,Stodolsky} for the quantum point contact as a detector,
the difference in that case is $S_I=2eI_0 (1-{\cal T})$ 
where ${\cal T}$ is the transparency of the channel \cite{Lesovik} 
(while above we implicitly assumed  
${\cal T}\ll 1$ \cite{largeT}). 
        Notice that the decoherence rate (\ref{gam_d-m}) was derived 
in Refs.\ \cite{Gurvitz,Aleiner,Levinson,Stodolsky} without any account of
the information provided by the detector, implicitly assuming that
the measurement result is just ignored. Now let us study how
this additional information affects the double-dot density matrix.

        We start with the completely classical case when there is no 
tunneling between dots ($H=0$) and
the initial density matrix of the system does not have nondiagonal 
elements, $\sigma_{12}(0)=\sigma_{12}(t)=0$. 
We can assume that the electron is actually located in one of the 
dots, but we just do not know exactly in which one, and that is why we 
use probabilities $\sigma_{11}(0)$ and $\sigma_{22}(0)=1-\sigma_{11}(0)$. 
The detector output is the fluctuating current $I(t)$. 
The fluctuations grow when we examine $I(t)$
at smaller time scales, so some averaging in time (``low-pass
filtering'') is necessary. Let us always work at sufficiently low 
frequencies, 
$f\sim \tau^{-1}\ll S_I/e^2$.

        The probability $P$ 
to have a particular value for the current averaged over time $\tau$,  
$\langle I\rangle =\int_0^\tau I(t) dt$, 
 is given by the distribution 
        \begin{eqnarray}
P(\langle I\rangle , \tau)=\sigma_{11}(0) \, P_1 (\langle I\rangle ,\tau )+ 
\sigma_{22}(0) \, P_2(\langle I\rangle ,\tau ), 
        \label{prob} \\
P_i(\langle I\rangle , \tau ) = (2\pi D)^{-1/2} 
\exp \left[ -(\langle I\rangle -I_i)^2/2D \right] , 
        \label{Gauss}\end{eqnarray}
where $ D = S_I/2\tau$.
After the measurement during
time $\tau$ the observer acquires additional knowledge about the system
and should change the probabilities $\sigma_{ii}$ according to the standard
Bayes formula for a posteriori probability. Hence,
        \begin{eqnarray}
\sigma_{11} (\tau ) &&= \sigma_{11}(0) \exp [ 
-(\langle I\rangle -I_1)^2/2D ] 
        \nonumber \\
&&\, \times \left\{ \sigma_{11}(0)\exp [ -(\langle 
 I\rangle -I_1)^2/2D ] \right.  
        \nonumber \\
 && \,\,\,\, + \left. \sigma_{22}(0) \exp  [- (\langle I\rangle -I_2)^2/2D] 
\right\} ^{-1} ,
        \nonumber \\
\sigma_{22} (\tau ) &&= 1-\sigma_{11} (\tau ) .
        \label{s1-2}\end{eqnarray}

        Now let us assume that the initial state is fully coherent,
$\sigma_{12}(0)=\sqrt{\sigma_{11}(0)\sigma_{22}(0)}$ (while still 
$H=\epsilon =0$). Since the detector is sensitive only 
to the position of electron, the detector current 
will behave {\it exactly} the same way.
So, after  the  measurement  during  time  $\tau$   we   should 
assign the same values for $\sigma_{ii}(\tau )$  
as in Eq.\ (\ref{s1-2}). Then the upper bound for the
nondiagonal element  is  
        \begin{equation}
\mbox{Re} \, \sigma_{12} (\tau ) \leq |\sigma_{12}(\tau )| \leq 
\sqrt{ \sigma_{11}(\tau ) \sigma_{22}(\tau )}. 
        \label{u-b}\end{equation}

        If the actual measurement result is disregarded, 
then the upper bound for $\sigma_{12}$ can be calculated 
using the probability distribution of different outcomes given by
Eq.\ (\ref{prob}) and the upper bound (\ref{u-b}) for each
realization: 
        \begin{eqnarray}
\langle \mbox{Re} \, \sigma_{12}(\tau )\rangle \leq 
\int \sqrt{\sigma_{11}(\tau )\sigma_{22}(\tau )} \, 
        P(\langle I\rangle ,\tau ) \, d\langle I\rangle 
        \nonumber \\
= \sqrt{\sigma_{11}(0)\sigma_{22}(0)} \, 
\exp [ -(\Delta I)^{2}\tau /4S_I ] .
        \end{eqnarray}
This upper bound exactly coincides with the result
given by decoherence approach (\ref{gam_d-m}). This fact
forces us to accept somewhat surprising statement
that Eq.\ (\ref{u-b}) gives not only the upper bound, but 
the true value of the nondiagonal 
matrix element, i.e.\ the pure state remains pure  
after the measurement (no decoherence occurs) if we know the
measurement result.
        
        Simultaneously, we proved that the point contact detector 
considered in Refs.\ \cite{Gurvitz,Aleiner,Levinson} causes the slowest
possible decoherence of the measured system, 
 and hence represents an ideal detector in this sense. 
In contrast, the result 
of Ref.\ \cite{Shnirman} shows that a single-electron transistor with 
large tunnel resistances and biased by relatively large voltage
is not an ideal detector (the non-ideal detector has been
also considered in  Ref.\ \cite{Stodolsky}).

        If the initial state of the double-dot is not purely coherent, 
$|\sigma_{12}(0)|<\sqrt{\sigma_{11}(0)\sigma_{22}(0)}$, 
it can be treated as the statistical combination of purely 
coherent and purely incoherent states with the same
$\sigma_{11}(0)$ and $\sigma_{22}(0)$, then 
        \begin{equation}
\sigma_{12}(\tau )= \sigma_{12}(0) \,  
[ \sigma_{11}(\tau ) \sigma_{22}(\tau )  /
 \sigma_{11}(0) \sigma_{22}(0)] ^{1/2}. 
        \label{s_12-g}\end{equation}
Eq.\ (\ref{s_12-g}) together with Eq.\ (\ref{s1-2}) is the central
result of the present letter.

        Equations (\ref{prob})--(\ref{s1-2}) and (\ref{s_12-g})
can be used to simulate the detector output $I(t)$ and the corresponding 
evolution of the density matrix. For example, in the 
Monte-Carlo method we should first choose
the timestep $\tau$ satisfying inequalities $e^2/S_I \ll \tau
\ll S_I/(\Delta I)^2$ and draw a random number for 
$\langle I\rangle$ according to the distribution (\ref{prob}). 
Then we update $\sigma_{11}(t)$ and $\sigma_{22}(t)$ using
this value of $\langle I\rangle$ and repeat the procedure many
times (the distribution for the current averaged over 
the interval $\Delta t=\tau$ 
is new every timestep because of changing $\sigma_{ii}(t)$ which
are used in Eq.\ (\ref{prob})).
The nondiagonal matrix element can be calculated at any time
using Eq.\ (\ref{s_12-g}).

        This Monte-Carlo procedure is equivalent to the following 
nonlinear Langevin-type equation for the density matrix evolution 
(equation for $\sigma_{11}$ is sufficient):
        \begin{eqnarray}
{\dot \sigma}_{11}={\cal R}&&,  \,\,\, \, {\cal R} =
-\sigma_{11}\sigma_{22}\, \frac{2\Delta I}{S_I} 
\left[ I(t)-I_0 \right] 
        \label{Rnew} \\ 
&&= -\sigma_{11}\sigma_{22}\, \frac{2\Delta I}{S_I}
\left[\frac{\sigma_{22}-\sigma_{11}}{2}\, \Delta I +\xi (t)  \right] ,
        \label{s_11-evol}\end{eqnarray}
where the random process $\xi (t)$ has zero average and the
low frequency spectral density $S_\xi = S_I$.
        The second expression for $\cal R$ allows to simulate the
measurement while the first one can be used to calculate the 
density matrix for given $I(t)$ (that can be done easier 
using Eq.\ (\ref{s1-2})).
        Notice that Eq.\ (\ref{s_11-evol}) is closely connected with the 
Quantum State Diffusion approach of Refs.\ 
\cite{Gisin,Carmichael,Plenio}. 

        Figure \ref{fig1} shows a particular result of the Monte-Carlo
simulation for the symmetric initial state, $\sigma_{11}(0)=
\sigma_{22}(0)=1/2$. Thick line shows the random evolution 
of $\sigma_{11}(t)$. Equation (\ref{s_11-evol}) describes the gradual 
localization in one of the dots (first dot in case of Fig.\ \ref{fig1}).
Let us define the typical localization time as
$\tau_{loc}=2S_I/(\Delta I)^2 $ (we choose the exponential factor
at $\sigma_{11}=\sigma_{22}=1/2$). Then it is exactly equal to the time
$\tau_{dis}=2S_I/(\Delta I)^2$ necessary for the observer to distinguish
between two states (defined as the relative shift of two Gaussians
(\ref{Gauss}) by two 
standard deviations), and $\tau_{loc}=\tau_d/2$ where $\tau_d=\Gamma_d^{-1}$.
It is easy to prove that the probability of final localization 
in the first dot is equal to $\sigma_{11}(0)$,   
because $\sigma_{ii}(\tau )$ averaged over realizations is conserved.

        The detector current $I(t)$ basically follows the evolution of
$\sigma_{ii}(t)$ but also contains the noise which
depends on the bandwidth. The dashed line in Fig.\ 
\ref{fig1} shows the current averaged over the ``running window''
with the duration $\Delta t= S_I/(\Delta I)^2$ while the thin solid
line is current $\langle I\rangle$ averaged starting from $t=0$. 
        Notice that our result for $I(t)$ directly contradicts the 
point of view presented in Ref.\ \cite{Stodolsky}.

        Now let us consider the general case of the double-dot
with non-zero tunneling $H$. If the frequency $\Omega$ of ``internal''
oscillations is sufficiently low, 
$ \Omega = (4H^2+\epsilon^2)^{1/2}/\hbar \ll S_I/e^2 $,
we can use the same formalism just adding the slow 
evolution due to finite $H$ (the product $\Omega \tau_{loc}$
is {\it arbitrary}). 
The particular realization can be
either simulated by Monte-Carlo procedure similar to that outlined 
above [now update of $\sigma_{12}(t)$ using Eq.\ (\ref{s_12-g}) should
be necessarily done at each timestep] 
 or equivalently described by the coupled
Langevin equations 
        \begin{eqnarray}
{\dot \sigma}_{11} &=& -{\dot \sigma}_{22}= (-2H/\hbar) \, 
        \mbox{Im}(\sigma_{12}) +{\cal R},
        \label{s11-g}   \\
{\dot \sigma}_{12} &=& \frac{i\epsilon }{\hbar} \sigma_{12} 
+\frac{iH}{\hbar}(\sigma_{11}-\sigma_{22}) +
\frac{\sigma_{22}-\sigma_{11}}{2\sigma_{11}\sigma_{22}}\, {\cal R} \sigma_{12} 
        \nonumber \\
&& -\gamma_d\sigma_{12},
        \label{s12-g}\end{eqnarray}
where $\gamma_d =0$ for an ideal detector (see below). 

        Figure \ref{fig2} shows particular results of the Monte-Carlo
simulations for the double-dot with $\epsilon =H$ and different
strength of the interaction with an ideal detector. The electron is initially
located in the first dot, $\sigma_{11}(0)=1$. The dashed line shows
the evolution of $\sigma_{11}$ with no detector. Notice that because
of $\epsilon \neq 0$, the initial asymmetry of the electron location 
remains in this case for infinite time. When the interaction with detector, 
${\cal C} =\hbar(\Delta I)^2/S_IH$, is relatively
small (top solid line), the evolution of $\sigma_{11}$ is close to
that without the detector. However, the electron gradually ``forgets''
the initial asymmetry and the evolution can be described as the slow
variation of the phase and amplitude of oscillations (recall that
the wavefunction remains pure). 

        When the coupling with the detector increases, the 
evolution significantly changes (middle and bottom curves in Fig.\
\ref{fig2}). First, the transition between dots slows down
(Quantum Zeno effect).
Second, while the frequency of transitions decreases with increasing
interaction with detector (at sufficiently strong coupling), 
the time of a transition decreases,
so eventually we can say about uncorrelated ``quantum jumps'' 
between states. 

        In a regime of small coupling with detector, ${\cal C}\ll 1$, 
the detector output is too noisy to follow the
evolution of $\sigma_{ii}$ and, correspondingly, only slightly 
affects the oscillations. On contrary,
when ${\cal C}\gg 1$ the detector accurately informs about the
position of electron and simultaneously
destroys the oscillations.  

        Equations (\ref{s11-g})--(\ref{s12-g}) can be generalized
for a nonideal detector, $\Gamma_d > (\Delta I)^2/4S_I$ 
(as in Refs.\ \cite{Stodolsky,Shnirman}), 
which gives an observer less information than possible in principle. 
Let us model it 
as two ideal detectors ``in parallel'' with unaccessible 
output of the second detector.
Then the information loss can be represented
by the extra decoherence term $-\gamma_d \sigma_{12}$ in Eq.\
(\ref{s12-g}) where $\gamma_d=\Gamma_d-(\Delta I)^2/4S_I$. 
The limiting case of a nonideal detector is the detector 
with no output (just an environment, $\Delta I=0$) or with disregarded output.
Then the evolution equations (\ref{s11-g})--(\ref{s12-g}) reduce to 
the standard decoherence approach. 

        For nonideal detector it is meaningful to keep our old 
definition of the localization time, $\tau_{loc}=\tau_{dis}=
2S_I/(\Delta I)^2$ while $\tau_d<2\tau_{loc}$. 
So, we consider localization time not as a real physical quantity  
(that is meaningless because observer cannot check it) but
as a quantity related to observer's information. Similarly,
the effective decoherence time is defined as $\tau_d'=\gamma_d^{-1}$.

Equations (\ref{s11-g})--(\ref{s12-g}) with the term $\cal R$ given
by Eq.\ (\ref{Rnew}) can be used to obtain  
the evolution of the density matrix in an experiment provided
the known detector output $I(t)$ 
and initial condition $\sigma_{ij}(0)$.
        Notice that even if the initial state is completely
random, $\sigma_{11}=\sigma_{22}=1/2$, $\sigma_{12}=0$, the
nondiagonal matrix element appears during the measurement 
due to acquired information, so that sufficiently long 
observation with an ideal detector leads to almost pure
wavefunction. Such a purification of the density matrix
described by Eqs.\ (\ref{s11-g})--(\ref{s12-g}) is analogous to
the localization at $H=0$. 

        Let us briefly discuss the philosophical aspect of the developed 
formalism. The statement that the pure wavefunction remains pure
during the continuous  measurement by an ideal detector (with  known  
result)  may  seem surprising 
at first, however, we easily recognize the direct analogy
to the ``orthodox'' situation of a ``sharp'' measurement (the wavefunction
is pure after the ``collapse''). Another important point
is that the density matrix is in some sense observer-dependent.
For example, if two
observers have different level of access to the detector information
(as, for example, in the model of nonideal detector considered above),
then the density matrix for them will be different. Nevertheless,
the observer with less information still can use his density matrix
for all purposes, just his predictions will be less accurate.
So, instead of the ``actual'' density matrix, 
it is better to discuss only the ``accessible'' density matrix,
that is fully consistent with the ``orthodox'' (Copenhagen) point of 
view.

        If the knowledge of the detector output is not used in the 
experiment, the decoherence approach is suitable. 
On contrary, one can devise 
an experiment in which the subsequent system evolution depends on
the preceding measurement result; then the proper description
is given by Eqs.\ (\ref{s11-g})--(\ref{s12-g}). 

For example, let us consider the double-dot with $H=0$
and fully coherent symmetric initial state. According to our formalism,
after the measurement by an ideal detector during time $\tau$ 
(most interesting
case is $\tau \alt \tau_{loc}$) the wavefunction remains pure
but becomes asymmetric (Eqs.\ (\ref{s1-2}) 
and (\ref{s_12-g})). 
To prove this, for example,
an experimentalist should switch off 
the detector at $t=\tau $,  
reduce the barrier between the dots (create finite $H$), 
and create the energy difference 
$\epsilon = [(1-4|\sigma_{12}|^2)^{1/2}-1] 
H \mbox{Re}\sigma_{12}/ |\sigma_{12}|^{2}$; 
then after the time period
$\Delta t =[\pi-\arcsin (\mbox{Im} \sigma_{12} \, \hbar\Omega/H)]/\Omega$
the electron will be moved to the
first dot with the probability equal to unity, 
that can be checked by the detector switched on again.
Alternatively, using the knowledge of $\sigma_{ij}(\tau )$ 
an experimentalist can exactly prepare the ground state of the coupled
double-dot system and check it, for example, by the photon
absorption.
        Another experimental idea is to start with completely random
state of the double-dot with finite $H$ and then gradually
(most interesting case is $\Omega \tau_{loc}\alt 1$)
obtain almost pure wavefunction using the detector output $I(t)$
and Eqs.\ (\ref{s11-g})--(\ref{s12-g}). The final test of
the wavefunction is similar to that considered above. 

An experiment of this kind can verify the formalism 
developed in the present letter. While such an experiment is
still a challenge for the present-day technology, we 
hope that it can be realized in the near future.

        In conclusion, we developed a simple formalism 
for the evolution of double-dot density matrix with account
of the result of the continuous measurement by weakly
coupled (weakly responding) point contact. The formalism 
is suitable for any two-level system measured by  
weakly coupled detector.  

        The author thanks S. A. Gurvitz, D. V. Averin, and K. K. Likharev 
for fruitful discussions. The work was supported in 
part by French MENRT (PAST), Russian RFBR, and 
Russian Program on Nanoelectronics.

\begin{figure}
\caption{Thick line:  particular Monte-Carlo realization 
of $\sigma_{11}$ evolution in time during the measurement of uncoupled
dots, $H=0$. 
Initial state is symmetric, $\sigma_{11}(0)=\sigma_{22}(0)=1/2$,
while the measurement leads to gradual localization. 
Initially pure wavefunction remains pure at any time $t$. 
Thin line shows the corresponding detector current 
$\langle I \rangle$ averaged over the whole time interval starting 
from $t=0$ while the dashed line is the current averaged over
the running window with duration $S_I/(\Delta I)^2$.
 }
\label{fig1}\end{figure}

\begin{figure}
\caption{
Random evolution of $\sigma_{11}$ (particular Monte-Carlo realizations)
for asymmetric double-dot, $\epsilon =H$, with the electron initially
in the first dot, $\sigma_{11}(0)=1$, for different 
strength of coupling with detector: 
${\cal C}=\hbar (\Delta I)^2/S_IH=$0.3, 3, and 30
from top to bottom. Dashed line represents ${\cal C}=0$ (unmeasured
double-dot). Increasing coupling with detector destroys the quantum
oscillations (while wavefunction remains pure at any $t$), slows down
the transitions between states (Quantum Zeno effect), and for 
${\cal C}\gg 1$ leads to uncorrelated jumps between well localized
states.
 }
\label{fig2}\end{figure}


\begin{references} 

\bibitem{Wheeler} {\it Quantum Theory of Measurement}, ed. by
        J. A. Wheeler and W. H. Zurek (Princeton Univ. Press, 
        Princeton, NJ, 1983).
        
\bibitem{Zurek} W. H. Zurek, Phys. Today, {\bf 44} (10), 36 (1991).

\bibitem{Gisin} N. Gisin, Phys. Rev. Lett. {\bf 19}, 1657 (1984).

\bibitem{Carmichael} H. J. Carmichael, {\it An open system approach
        to quantum optics}, Lecture notes in physics (Springer, Berlin, 
        1993).

\bibitem{Plenio} M. B. Plenio and P. L. Knight, Rev. Mod. Phys. 
        {\bf 70}, 101 (1998).

\bibitem{q-comp} C. Bennett, Phys. Today, Oct. 1995, 24 (1995).

\bibitem{Buks} E. Buks, R. Schuster, M. Heiblum, D. Mahalu,
        and V. Umansky, Nature {\bf 391}, 871 (1998).

\bibitem{Gurvitz} S. A. Gurvitz, Phys. Rev. B {\bf 56}, 15215 (1997);
        quant-ph/9806050.

\bibitem{Aleiner} I. L. Aleiner, N. S. Wingreen, and Y. Meir,
        Phys. Rev. Lett. {\bf 79}, 3740 (1997).

\bibitem{Levinson} Y. Levinson, Europhys. Lett. {\bf 39}, 299 (1997).

\bibitem{Stodolsky} L. Stodolsky, quant-ph/9805081.

\bibitem{bigDI} If $\Delta I \sim I_0$, the evolution depends on the
        interaction between the detector and the next measuring stage.

\bibitem{Lesovik} G. B. Lesovik, JETP Lett. {\bf 49}, 591 (1989). 

\bibitem{largeT} In the case $1-{\cal T}\ll 1$, Eq.\ (\protect\ref{weak})
        should be replaced by $|\Delta I| \ll (1-{\cal T}) I_0
        \sim S_I/e$.

\bibitem{Shnirman} A. Shnirman and G. Sch\"on, 
        Phys. Rev. B {\bf 57}, 15400 (1998). 

 
\end{references}
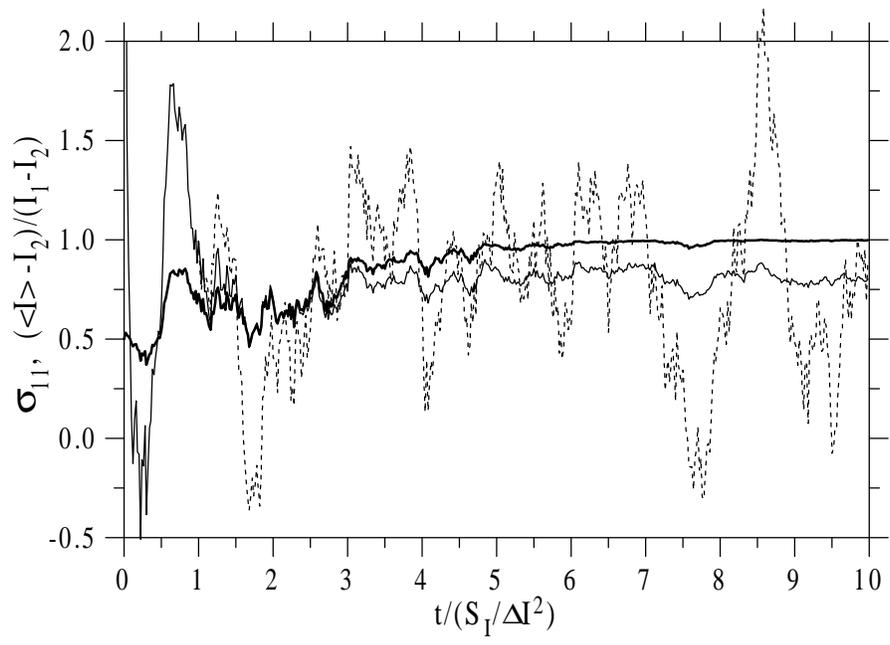
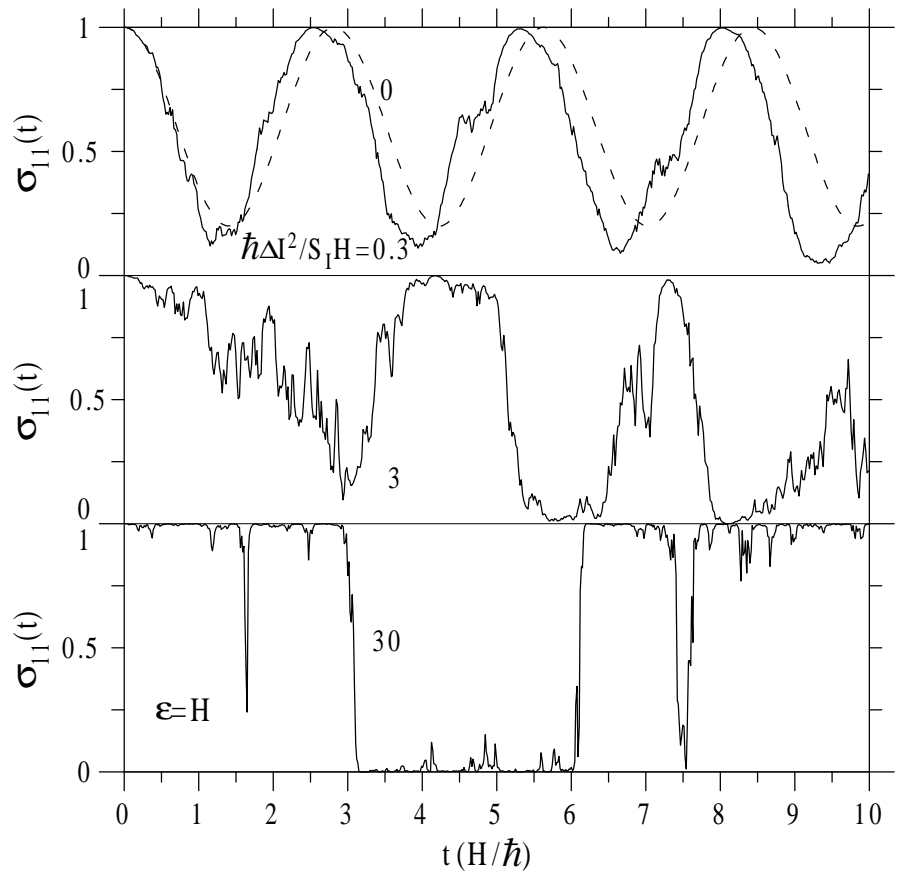
\end{document}